\documentclass[12pt]{article}
\usepackage{amsmath}
\usepackage{graphicx}
\usepackage{enumerate}
\usepackage{natbib}
\setcitestyle{aysep={}}
\usepackage{url} % not crucial - just used below for the URL 
\usepackage{booktabs}
\usepackage{enumitem} 
\usepackage{listings}

\usepackage{lineno,hyperref}
\usepackage{hyperref}       % hyperlinks
\usepackage{booktabs}     
\usepackage{multirow}  % professional-quality tables
\usepackage{amsfonts,bm}       % blackboard math symbols
\usepackage{nicefrac}       % compact symbols for 1/2, etc.
\usepackage{microtype}      % microtypography
\usepackage{lipsum}
\usepackage{float}
\usepackage[ruled,vlined]{algorithm2e}

\usepackage{graphicx}
\usepackage{subcaption}

\newcommand{\blind}{0}

\begin{document}

\def\spacingset#1{\renewcommand{\baselinestretch}%
{#1}\small\normalsize} \spacingset{1}

%%%%%%%%%%%%%%%%%%%%%%%%%%%%%%%%%%%%%%%%%%%%%%%%%%%%%%%%%%%%%%%%%%%%%%%%%%%%%%

\if0\blind
{
  \title{\bf Multiple Imputation Through XGBoost}
  \author{Yongshi Deng \\
    Department of Statistics, University of Auckland\\
    and \\
    Thomas Lumley\\
   Department of Statistics, University of Auckland}
  \maketitle
} \fi

\if1\blind
{
  \bigskip
  \bigskip
  \bigskip
  \begin{center}
    {\LARGE\bf MI through XGBoost}
\end{center}
  \medskip
} \fi

\bigskip
\begin{abstract}
The use of multiple imputation (MI) is becoming increasingly popular for addressing missing data. Although some conventional MI approaches have been well studied and have shown empirical validity, they have limitations when processing large datasets with complex data structures. Their imputation performances usually rely on the proper specification of imputation models, and this requires expert knowledge of the inherent relations among variables. Moreover, these standard approaches tend to be computationally inefficient for medium and large datasets. In this paper, we propose a scalable MI framework \textbf{mixgb}, which is based on XGBoost, subsampling, and predictive mean matching. Our approach leverages the power of XGBoost, a fast implementation of gradient boosted trees, to automatically capture interactions and nonlinear relations while achieving high computational efficiency. In addition, we incorporate subsampling and predictive mean matching to reduce bias and to better account for appropriate imputation variability. The proposed framework is implemented in an \texttt{R} package \textbf{mixgb}. Supplementary materials for this article are available online.

\end{abstract}

\noindent%

\vfill

\newpage
\spacingset{1.5} 
\section{Introduction}\label{sec:1}

Multiple imputation (MI), first introduced by \citet{Rubin1978}, has received increasing recognition in dealing with missing data because it can reduce bias and represent the uncertainty of missing values. Given an incomplete dataset, each missing value is replaced by a set of $M>1$ plausible values instead of a single value. Analyses can then be performed separately on $M$ complete imputed datasets, and the results can be combined to yield valid inference. \cite{Rubin1987} argued that with proper multiple imputation, the pooled estimates and variance would be statistically valid. A number of frameworks have been developed to implement MI. However, few automated procedures for large-scale imputation have been devised.

A major flaw of traditional MI implementations is that they fail to automatically capture complex relations among variables. If such relations exist, popular software packages such as \textbf{mi} \citep{Su2011} and \textbf{mice} with default settings \citep{Buuren2011} will produce unsatisfactory results, unless users manually specify any potential nonlinear effects in the imputation model for each incomplete variable. Indeed, \cite{Buuren2011} indicated that it is crucial to include all interactions of interest in the imputation model in order to achieve optimal results. However, it appears that researchers often use \textbf{mice} in an automated manner \citep[e.g.][]{Wendt2021,Awada2021}. Tree-based algorithms provide a potential solution to solve this problem. \cite{Stekhoven2012} proposed a nonparametric method for imputing missing values based on random forests \citep{Breiman2001} and implemented it in an \texttt{R}  called \textbf{missForest}. \cite{Doove2014} implemented classification and regression trees (CART) \citep{Breiman1984} and random forests within the \textbf{mice} framework (\texttt{mice-cart} and \texttt{mice-rf}) and showed that they can better preserve nonlinear effects, compared to the standard implementation of \textbf{mice}. 

Another drawback of existing MI frameworks is the excessive computation time for large datasets. Recent advances in technology have made the collection and analysis of large datasets feasible. However, current MI methods, including those that can capture complex data structures such as \texttt{mice-cart} and \texttt{mice-rf}, are more suited for small datasets and often struggle with medium and large datasets. This problem can become unmanageable when there is a moderate number of missing values present across many variables. Recently, \textbf{mice} has included \textbf{ranger} \citep{Wright2017}, a fast implementation of random forests, as one of its imputation methods. However, its imputation performance and computational time have not been investigated.

The purpose of this paper is to present a fast and automated MI procedure \textbf{mixgb}, which is based on XGBoost \citep{Chen2016}, subsampling, and predictive mean matching \citep{Little1988}, with a focus on yielding statistically valid results. XGBoost, a fast tree boosting algorithm, has been a frequent winner in Kaggle data competitions \citep{Chen2016} and has gained immense popularity due to its speed and accuracy. XGBoost's ability to efficiently capture complex structures of large datasets means it has great potential for automated MI. With subsampling and predictive mean matching, our proposed method can better incorporate the variability of missing data and enhance imputation quality.

This paper is structured as follows. Section \ref{sec:2} describes the proposed MI through XGBoost framework (\textbf{mixgb}) in detail. Section \ref{sec:3.1} provides an overview of the simulation study that we used to evaluate imputation performance. The evaluation criteria for MI implementations are presented in Section \ref{sec:3.2}. Simulation results are given in Section \ref{sec:3.3}, demonstrating the imputation quality of \textbf{mixgb}. Section \ref{sec:4} demonstrates the advantage of \textbf{mixgb} over other implementations in terms of computational efficiency, and Section \ref{sec:5} provides an example using a real dataset. Finally, discussions based on empirical results are presented in Section \ref{sec:6}. We have implemented the proposed method in our \texttt{R}  \textbf{mixgb} \citep{Deng2023}, which is available at \url{https://github.com/agnesdeng/mixgb} and on CRAN.

\section{Framework}\label{sec:2}
Using XGBoost to impute missing data has attracted growing interest in recent years. However, previous work has focused on the prediction accuracy of missing values without adequately accounting for the uncertainty of missing data, resulting in an underestimation of imputation variability \citep[e.g.,][]{Zhang2019}. To address this problem, we propose using XGBoost with subsampling and predictive mean matching (PMM) for MI. 

Our \texttt{R} package \textbf{mixgb} offers settings that allow users to use subsampling with different ratios and choose the type of PMM. In this paper, we use \texttt{mixgb} to denote using XGBoost with only PMM, and \texttt{mixgb-sub} to denote using XGBoost with PMM and subsampling. Unlike the \texttt{R} package \textbf{mice}, our imputation framework is noniterative. However, users can set the number of iterations in our package \textbf{mixgb} to be greater than one. In Section \ref{sec:3}, the imputation performance of both \texttt{mixgb} and \texttt{mixgb-sub} was examined using a single iteration, whereas \textbf{mice} was evaluated with five iterations.

Our framework \texttt{mixgb-sub} works as follows: Given a dataset $Y_{\text{raw}}$ with $p$ incomplete variables, we first sort the variables by the number of missing values in ascending order. We then conduct an initial imputation of the missing values to obtain a complete sorted dataset $Y$. We let $Y_i^{\text{obs}}$ and $Y_i^{\text{mis}}$ be the observed values and the imputed (originally missing) values for an incomplete variable $Y_i$. We also let $Y_{-i}^{\text{obs}}$ denote the corresponding data in all variables other than $Y_i$ for entries where $Y_i$ was observed. Similarly, $Y_{-i}^{\text{mis}}$ denotes the data for variables other than $Y_i$ for entries where $Y_i$ was originally missing before the initial imputation.

We use XGBoost models with subsampling to better account for the uncertainty of missing values. This is analogous to sampling parameters of a parametric model from their posterior distributions. When the subsampling ratio is less than one, a different subset of the data is used in each boosting round. Therefore, the set of data $Y^*$ used in each of the $M$ imputations will differ. For each incomplete variable, we fit an XGBoost model with subsampling and use it to obtain predictions for $Y_i^{\text{mis}}$, which we denote by $\widetilde{Y_i}^{*\text{mis}}$.

However, for continuous data, simply imputing missing values with point predictions $\widetilde{Y_i}^{*\text{mis}}$ is prone to underestimating imputation variability. This problem can be mitigated by PMM, which was first proposed by \cite{Rubin1986} and extended to MI by \cite{Little1988}. PMM works by matching the predicted value of each missing entry to a set of $K$ donors that have the closest predicted values among the observed entries. One donor is then randomly selected, and its observed value is used to impute the missing value. We write $\widehat{\beta}_i$ for the estimates of model  $f:Y_i^{\text{obs}}\sim Y_{-i}^{\text{obs}}$ using the whole data and $\widetilde{\beta}_i^*$ for the estimates of model  $f^*:Y_i^{*\text{obs}}\sim Y_{-i}^{*\text{obs}}$ using subsampling for each imputation. Four types of PMM are summarized in Table \ref{tab:pmm} based on \Citet{Buuren2018}.

\begin{table}[H]
		\centering
	\resizebox{\textwidth}{!}{%
	\begin{tabular}{c|c|c|c}
		\hline
		\textbf{PMM} &\multicolumn{1}{c|}{\textbf{\begin{tabular}[c]{@{}c@{}}Donors' \\ predicted values\end{tabular}}}&\multicolumn{1}{c|}{\textbf{\begin{tabular}[c]{@{}c@{}}Recipients' \\ predicted values\end{tabular}}}    & \textbf{Differences} \\ \hline
		Type 0                 & $\widehat{Y}_i^{\text{obs}}=Y_{-i}^{\text{obs}} \widehat{\beta}_i$ &       $\widehat{Y}_i^{\text{mis}}=Y_{-i}^{\text{mis}}\widehat{\beta}_i$               &     $\widehat{\beta}_i$ are the same across $M$ imputations     \\
		Type 1                 &  $\widehat{Y}_i^{\text{obs}}= Y_{-i}^{\text{obs}} \widehat{\beta}_i$   &    $\widetilde{Y}_i^{*\text{mis}}$=$Y_{-i}^{\text{mis}} \widetilde{\beta}_i^*$                  &   donors use $\widehat{\beta}_i$ and recipients use $\widetilde{\beta}_i^*$ for each imputation    \\
		Type 2                  &  $\widetilde{Y}_i^{*\text{obs}}= Y_{-i}^{\text{obs}}\widetilde{\beta}_i^*$   &    $\widetilde{Y}_i^{*\text{mis}}$=$Y_{-i}^{\text{mis}} \widetilde{\beta}_i^*$                  &    donors and recipients both use $\widetilde{\beta}_i^*$ for each imputation  \\
		Type 3                 &         $\widetilde{Y}_i^{'\text{obs}}= Y_{-i}^{\text{obs}}\widetilde{\beta}_i^{'}$                               &  $\widetilde{Y}_i^{''\text{mis}}$=$Y_{-i}^{\text{mis}}  \widetilde{\beta}_i^{''}$                    & $\widetilde{\beta}_i^{'}$ and $\widetilde{\beta}_i^{''}$ are two different draws for each imputation     \\ \hline
	\end{tabular}
}
	\caption{Four types of predictive mean matching (PMM).} \label{tab:pmm}
\end{table}

We have implemented Type 0, Type 1 and Type 2 PMM in our \texttt{R} package \textbf{mixgb}. Type 0 is improper as it ignores the sampling variability in $\widehat{\beta}_i$, so we offer this option only for research purposes. Type 3 needs two distinct draws of estimates for each imputation, which requires training $2M$ imputation models. We consider it computationally inefficient, hence it is not included in the current version. For Type 1 PMM, the predicted values for the donors are obtained by fitting a model to the entire dataset, whereas the predicted values for the recipients are produced by fitting an imputation model with subsampling for each imputation. For Type 2 PMM, the same model is used to obtain the predicted values for the donors and recipients. By default, our framework uses Type 2 PMM for continuous data and no PMM for categorical data. The detailed algorithm \texttt{mixgb-sub} is described in Algorithm \ref{algo:mixgb-sub} and illustrated in Figure \ref{fig:algorithm}. Without subsampling, \texttt{mixgb} works in a similar way. However, the entire dataset is used for each imputation. Note that omitting subsampling will reduce the variability between imputations and the primary source of between-imputation variance will be due to the differences between the PMM donors. If the number of PMM donors is set to one and subsampling is not used, then each of the $M$ imputations will be identical.

\begin{figure}[H]
	\centering
	\includegraphics[width=\textwidth,keepaspectratio]{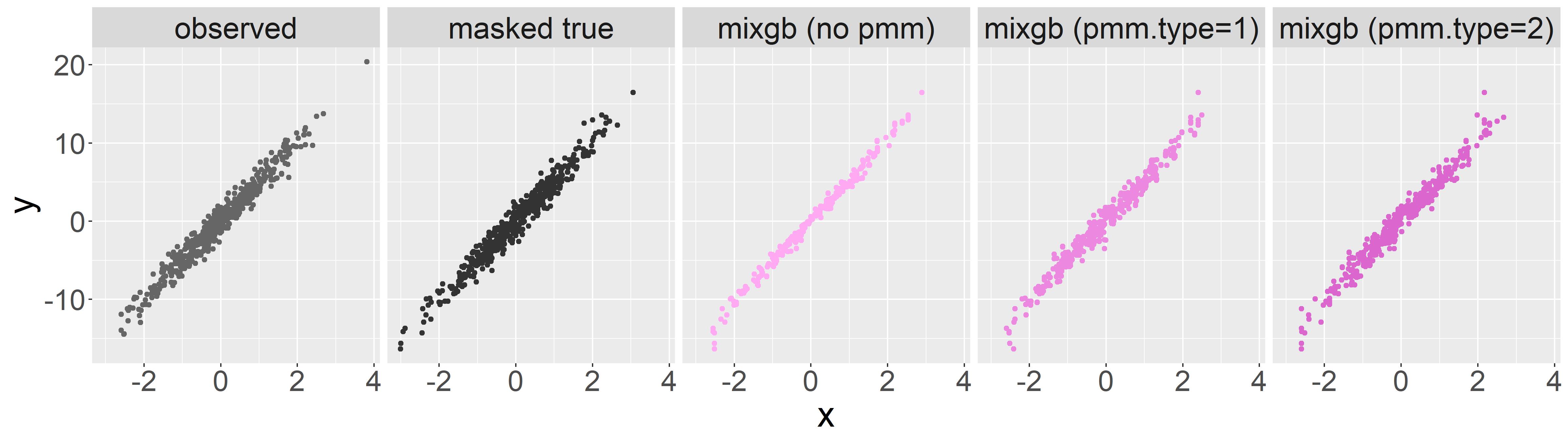}
	\caption{Scatter plots of $y$ versus $x$ for observed data, masked true data, imputed data using \texttt{mixgb} without PMM, imputed data using \texttt{mixgb} with Type 1 PMM, and imputed data using \texttt{mixgb} with Type 2 PMM. It demonstrates that without the use of PMM, the within-imputation variance of \texttt{mixgb} is considerably lower than the variance of both the observed and masked true data.}
	\label{fig:pmm}
\end{figure}

Figure \ref{fig:pmm} illustrates the need for using PMM for imputing continuous data with \texttt{mixgb}. We created a dataset of 1000 observations using $y_i=5x_i+\epsilon_i$, where $x_i\sim \text{Normal}(0,1) $ and $\epsilon_i \sim \text{Normal}(0,1)$ for $i=1, 2, ...., 1000$. We then generated 50\% missing data in the variable $x$ under the missing completely at random (MCAR) mechanism. We refer to simulated observations chosen to be missing as ``masked true''. As shown in Figure \ref{fig:pmm}, the variability within an imputed dataset using \texttt{mixgb} without PMM was considerably less than the variance of the observed data and the variance of the masked true data. In contrast, both Type 1 and Type 2 PMM had similar within-variance to the masked true data. 

\begin{figure}
\centering
	\begin{algorithm}[H]
	
		\small
		\SetArgSty{textnormal}
		\KwIn{
			a dataset $Y_{\text{raw}}$ with $p$ incomplete variables \;
			the number of imputations $M$ (by default, $M=5$)\;
			the number of iterations $T$ (by default, $T=1$)\;
			the type of PMM (By default, Type 2 for continuous data and no PMM for categorical data).
		}
		
		{\bf Initialization:} 
		sort variables by the number of missing values in ascending order\;
		make an initial guess for missing data and obtain a complete sorted dataset $Y_\text{init}$\;

		\uIf{pmm.type=1}{
			$Y\leftarrow Y_\text{init}$\\
			\For{$i=1$ \KwTo $p$}{
				fit an XGBoost model using the entire dataset $Y$ to obtain $\widehat{\beta}_i$\;
				$\widehat{Y}_i^{\text{obs}}\leftarrow$ using $\widehat{\beta}_i$ to get predicted values of $Y_i^{\text{obs}}$\;
				
			}
		}
		
		\For{$m=1$ \KwTo $M$} {
			$Y^{(1)}\leftarrow Y_\text{init}$\;			
			\For{$t=1$ \KwTo $T$}{
                  $Y^{(t)*}$: a subsample set from the whole dataset $Y^{(t)}$.\\
		
				\For{$i=1$ \KwTo $p$}{
					fit an XGBoost model with a subsample set of data $Y^{(t)*}$ to obtain $\widetilde{\beta}_i^*$\;
					$\widetilde{Y}_i^{*\text{mis}}\leftarrow$ using  $\widetilde{\beta}_i^*$ to get the predicted values of $Y_i^{\text{mis}}$\;
					
					\uIf{pmm.type=1}{
						update $Y^{(t)}$ with donors from $Y_i^{\text{obs}}$ by matching $\widetilde{Y}_i^{*\text{mis}}$ to $\widehat{Y}_i^{\text{obs}}$\;
					}
					\uElseIf{pmm.type=2}{
						$\widetilde{Y}_i^{*\text{obs}}\leftarrow$ using $\widetilde{\beta}_i^*$ to get predicted values of $Y_i^{\text{obs}}$\;
						update $Y^{(t)}$ with donors from $Y_i^{\text{obs}}$ by matching $\widetilde{Y}_i^{*\text{mis}}$ to $\widetilde{Y}_i^{*\text{obs}}$\;
					}
					\uElse{update $Y^{(t)}$ with $\widetilde{Y}_i^{*\text{mis}}$}
					
				}
				$Y^{(t+1)} \leftarrow Y^{(t)}$.
			}

			$Y_{\text{imp-m}}\leftarrow$ reorder the variables of $Y^{(T)}$ \;
		
			\KwRet{the $m^\text{th}$ imputed dataset $Y_{\text{imp-m}}$}
		}
				
		\KwOut{
			$M$ imputed datasets $Y_{\text{imp}}=(Y_{\text{imp-1}},Y_{\text{imp-2}},...,Y_{\text{imp-M}})$
		}
		
		\caption{MI Through XGBoost With PMM and Subsampling (\texttt{mixgb-sub})}
		\label{algo:mixgb-sub}
	\end{algorithm} 
\end{figure}

\begin{figure}[H]
	\centering
	\includegraphics[width=0.9\textwidth,keepaspectratio]{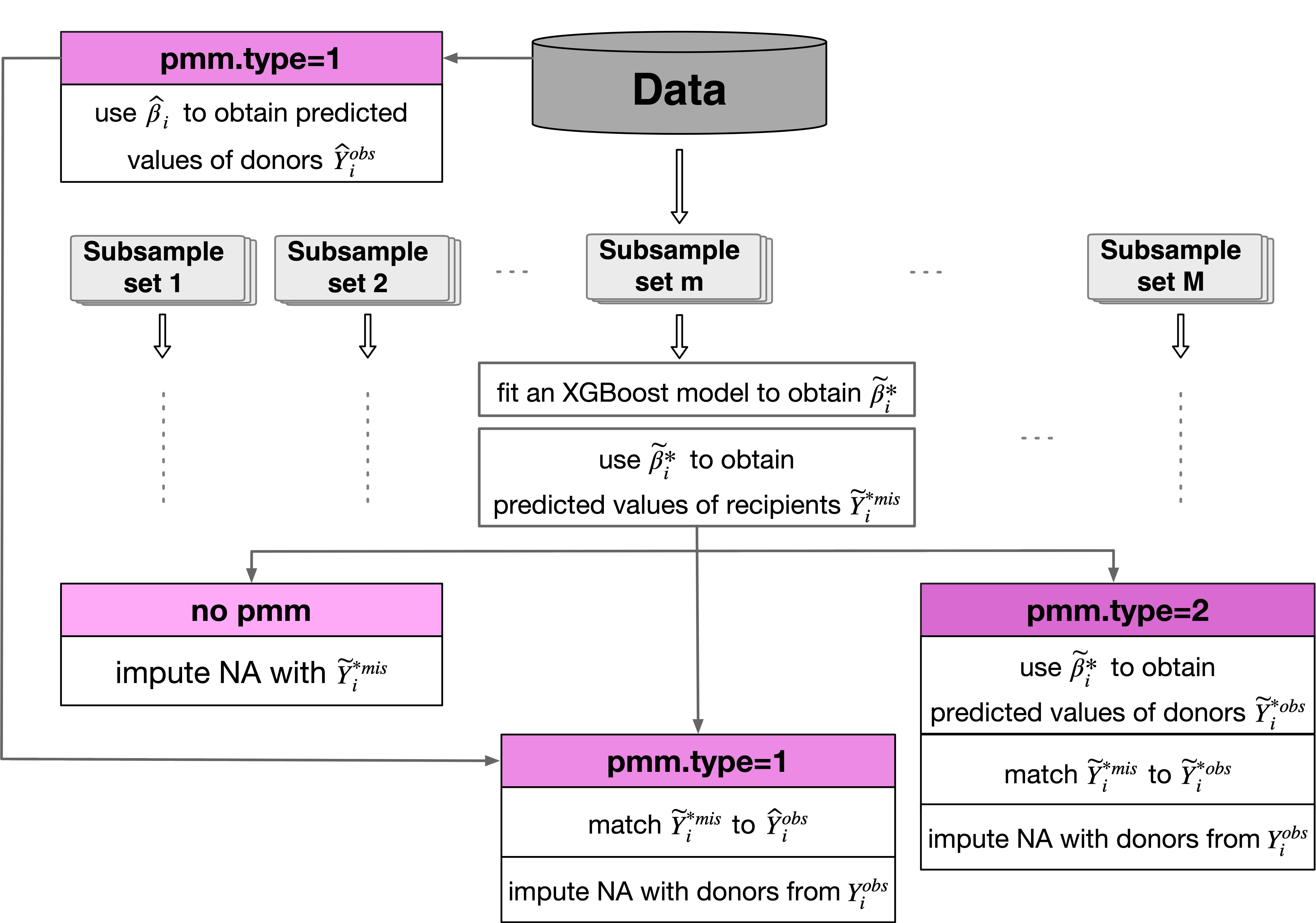}
	\caption{Visualization of Algorithm \ref{algo:mixgb-sub} for \texttt{mixgb-sub}.}
	\label{fig:algorithm}
\end{figure}

\section{Simulation Study}\label{sec:3}
\subsection{Overview} \label{sec:3.1}
We generated a single complete dataset using data generation model (\ref{mixgbsim}). For each simulation run, we generated missing data via a missing at random mechanism (MAR) and obtained $M=5$ imputed datasets through different MI methods.

\paragraph{Data Generation.}
A dataset with mixed-type variables was generated using the following data generation model for $i = 1, 2, ..., 10000$:
\begin{equation}\label{mixgbsim}
\begin{split}
Y_i &= \texttt{norm1}_i+\texttt{norm2}_i + \texttt{norm3}_i  +\texttt{norm5}_i+\texttt{norm7}_i\\
&+(\texttt{bin1}_i=1)-( \texttt{ord1}_i=1)-2\cdot( \texttt{ord1}_i=2)\\
&+\texttt{norm1}_i^2+\texttt{norm2}_i\times\texttt{norm3}_i-3\cdot \texttt{norm5}_i\times(\texttt{bin1}_i=1) \\
& -2\cdot \texttt{norm7}_i\times(\texttt{ord1}_i=1) + \texttt{norm7}_i\times(\texttt{ord1}_i=2)+\epsilon_i,
\end{split}
\end{equation}

where the $\epsilon_i$ was from a standard normal distribution. The variables $\texttt{norm1},\dots, \texttt{norm8}$ were drawn from standard normal distributions, $\texttt{bin1}$ was drawn from a binary distribution with $p=0.5$, and $\texttt{ord1}$ was drawn from a binomial distribution with $n=2$ and $p=0.5$. The distributions of $\texttt{norm1},\texttt{norm2},\texttt{norm3}$ and $\texttt{norm4}$ had pairwise correlations of $0.5$. The binary variable $\texttt{bin1}$ was correlated with both $\texttt{norm5}$ and $\texttt{norm6}$ with correlation $\rho\approx 0.55$, while $\texttt{norm5}$ and $\texttt{norm6}$ had correlation of $0.7$. The ordinal variable $\texttt{ord1}$ was correlated with both $\texttt{norm7}$ and $\texttt{norm8}$ with correlation $\rho\approx 0.65$, while $\texttt{norm7}$ and $\texttt{norm8}$ had correlation of $0.7$. Note that  $\texttt{norm4}$, $\texttt{norm6}$ and $\texttt{norm8}$ were included in the dataset as ancillary variables, but they were not used to generate $Y$.

\paragraph{Missing Data Mechanism.}
All variables, except for $Y$ and the three ancillary variables $\texttt{norm4}$, $\texttt{norm6} $ and $\texttt{norm8}$, were made missing via an MAR mechanism depending on $Y$ and one of the three ancillary variables. Let $Z_i= Y_i + \texttt{norm4}_i$. Suppose that $R_i=0$ when $\texttt{norm1}_i$ is missing. We let 
\[  P(R_i=0) =\left\{
\begin{array}{ll}
0.6 & \text{if }Z_i \text{ is in the top third of }Y+\texttt{norm4}, \\
0.1 & \text{if }Z_i \text{ is in the middle third of }Y+\texttt{norm4}, \\
0.6 & \text{if }Z_i \text{ is in the bottom third of }Y+\texttt{norm4}.
\end{array}
\right. \]
Similarly, missing values in $\texttt{norm2}$ and $\texttt{norm3}$ were generated using the same mechanism based on $Y+\texttt{norm4}$. Missing values in $\texttt{norm5}$ and $\texttt{bin1}$ were generated using the same method, except with $Y+\texttt{norm6}$ in place of $Y+\texttt{norm4}$. Missing values in $\texttt{norm7}$ and $\texttt{ord1}$ were similarly generated based on $Y+\texttt{norm8}$.

\paragraph{MI Methods.}
We assessed the imputation quality of the following implementations: default settings within \textbf{mice} (\texttt{mice-default}), classification and regression trees within \textbf{mice} (\texttt{mice-cart}), fast implementation of random forests within \textbf{mice} (\texttt{mice-ranger}), XGBoost with PMM (\texttt{mixgb}), and XGBoost with PMM and subsampling (\texttt{mixgb-sub}).

We used the default settings for methods from the \texttt{R} package \textbf{mice}, including the number of iterations \texttt{maxit = 5}, whereas the default number of iterations in \texttt{mixgb} and \texttt{mixgb-sub} is one. We believe that our method has good performance even when the number of iterations is smaller than that of \textbf{mice}. The performance of all methods was evaluated over 1000 simulation runs. Our evaluation was based on empirical bias, within-imputation variance, between-imputation variance and coverage.

\subsection{Evaluation Criteria}\label{sec:3.2}
Suppose $\beta$ is a $k\times 1$ column vector, representing the true coefficients of data generation model \ref{mixgbsim}. Let $\hat{\beta}$ be the estimate of $\beta$ for the full simulated dataset and let $U$ be estimated variance-covariance matrix of $\hat{\beta}$. To investigate the imputation performance of MI methods, we used the following criteria based on \cite{Rubin1987} and \cite{Brand2003}:

\paragraph{Empirical Bias.} For each simulation run, we obtained an MI combined estimate $\overline{\beta}_M$ for each imputation method, where $\overline{\beta}_M=\frac{1}{M}\sum_{m=1}^M \beta_m^*$ and $\beta_m^*$ is the estimate of the $m^\text{th}$ imputation. We are interested in the empirical bias $E_h[\overline{\beta}_M]- \widehat{\beta}$ , where $E_h$ denotes averaging over 1000 simulation runs.

\paragraph{Within-imputation Variance.} The within-variance component of the combined MI estimate is defined as $\overline{U}_M = \frac{1}{M}\sum_{m=1}^M U_m^*$, where $U_m^*$ is the variance-covariance matrix of $\beta_m^*$. We compared the average of within-imputaion variance over 1000 runs $\text{Var}_W:=E_h[\overline{U}_M]$ to $\text{Var}_W^{\text{target}}:=U$, the estimated variance-covariance of $\widehat{\beta}$.

\paragraph{Between-imputation Variance.} The between-imputation variance is given by $B_M=\frac{1}{M-1} \sum_{m=1}^M(\beta_m^*-\overline{\beta}_M) (\beta_m^*-\overline{\beta}_M)^t$. The average adjusted between-imputation variance over 1000 simulations is $\text{Var}_B:=(1+M^{-1})E_h[B_M]$, where the factor $(1+M^{-1})$ is used to adjust for $M$ being finite. We compared $\text{Var}_B$ to $\text{Var}_B^{\text{target}}:=\widehat{\text{Var}}(\overline{\beta}_M)$, the empirical variance of the $\overline{\beta}_M$ over 1000 simulation runs.

\paragraph{Coverage.} Coverage is the proportion of replications where the 95\% confidence interval around the combined MI estimate $\overline{\beta}_M$ contains $\widehat{\beta}$. The 95\% confidence interval for $\widehat{\beta}$ is calculated as $\overline{\beta}_M\pm t_{\nu,0.975}\sqrt{(1+M^{-1})B_M}$, where $t_{\nu,0.975}$ is the 97.5\% quantile of a Student-t distribution with the degrees of freedom $\nu$ as defined in \cite{Rubin1987}.

\subsection{Results}\label{sec:3.3}
\begin{figure}[H]
	\centering
	\includegraphics[width=\textwidth]{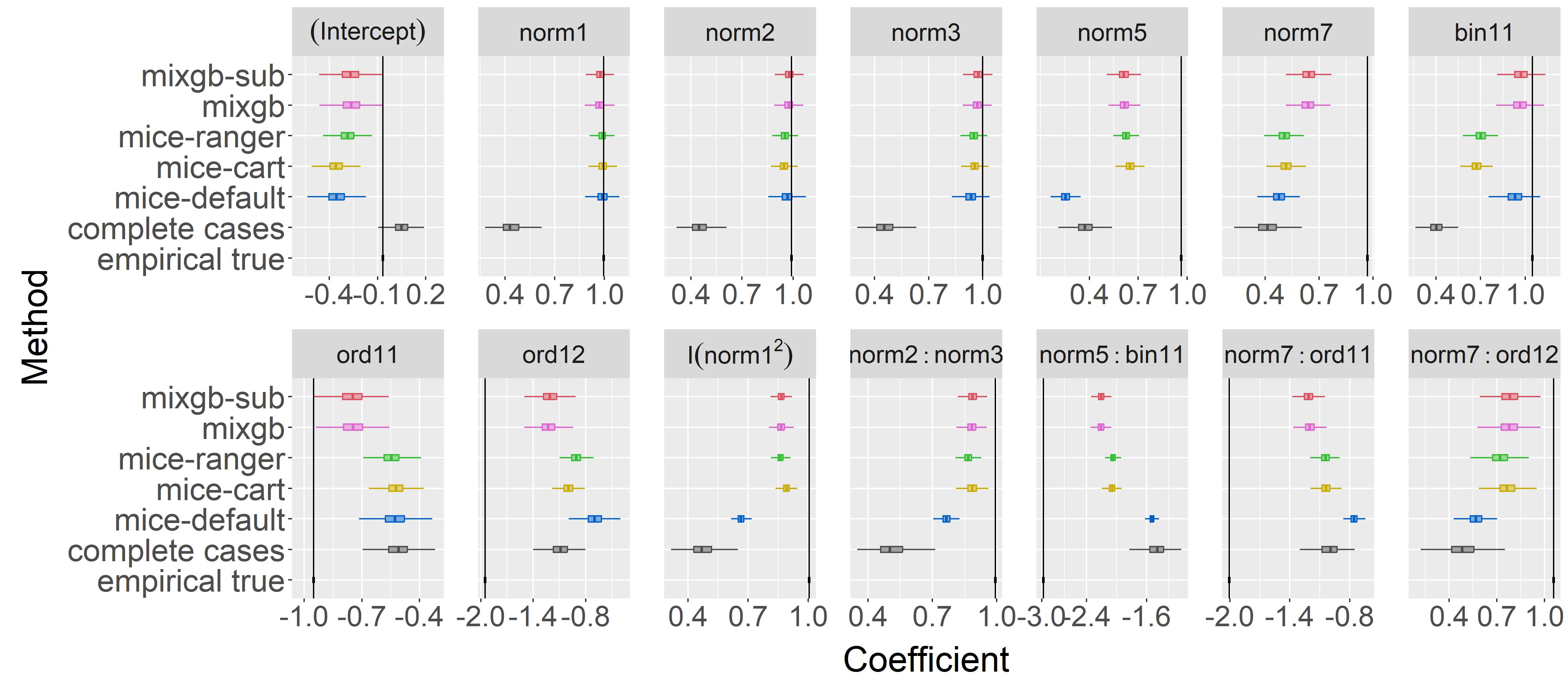}
	\caption{Visualization of the simulation results on coefficient estimates of MI methods and complete case analysis (1000 runs). Vertical solid lines represent the empirical true values. It shows that both \texttt{mixgb} and \texttt{mixgb-sub} obtained less biased estimates than \texttt{mice-default} and complete case analysis, and as good as or better than \texttt{mice-cart} and \texttt{mice-ranger}.} \label{fig:coefs}
\end{figure}

\begin{figure}[H]
	\centering
	\includegraphics[width=\textwidth]{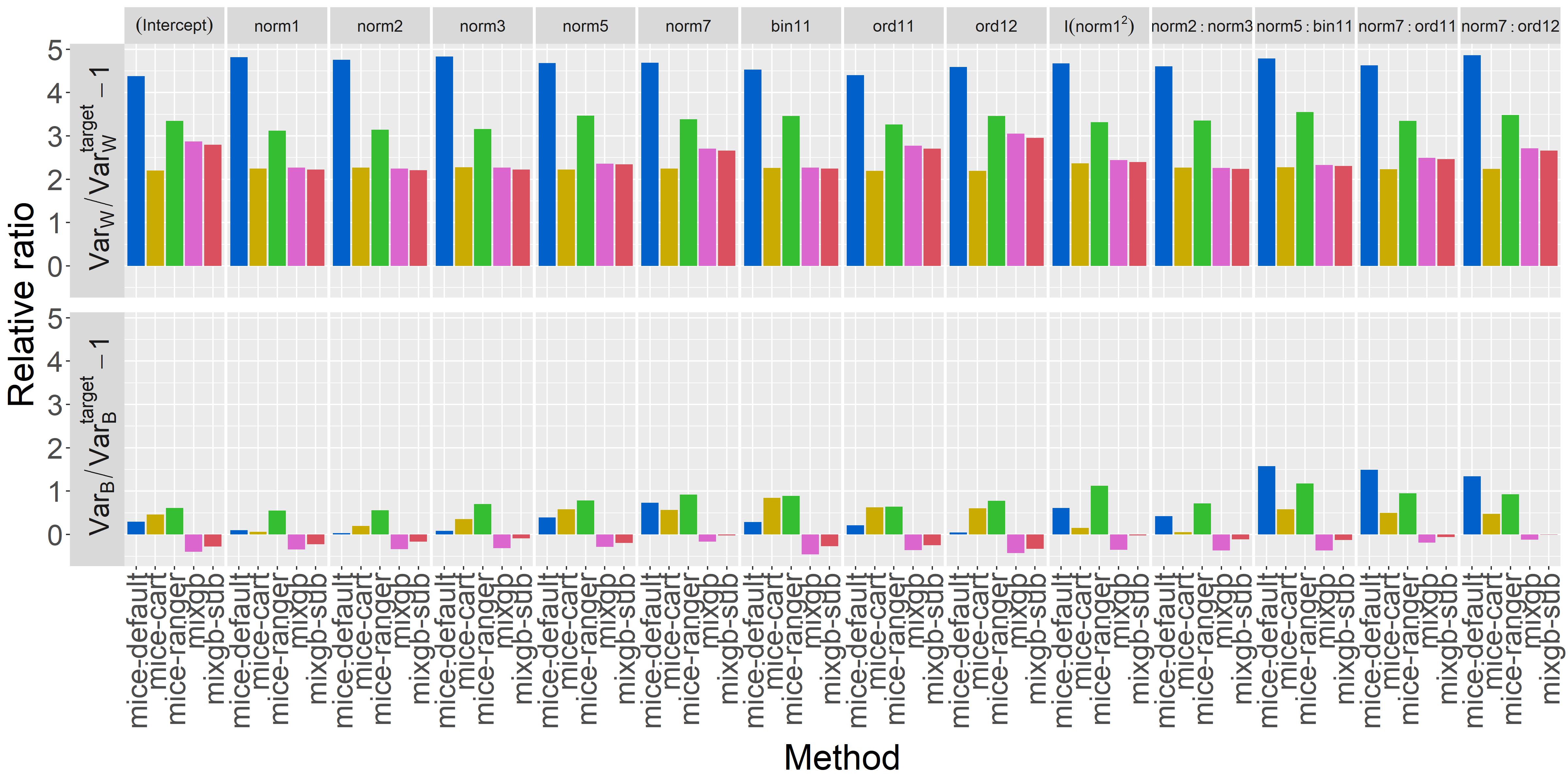}
	\caption{Relative ratio of imputation variance and target imputation variance of each estimate using MI methods over 1000 simultation runs. The top panel compares performance on within-imputation variability, and the bottom panel shows performance on between-imputation variability. Relative ratios that are closer to zero signify better estimates in imputation variance.}\label{fig:var}
\end{figure}

Simulation results over 1000 runs are summarized in Table \ref{tab:simtable}. We visualized the simulation results of biases and imputation variances in Figure \ref{fig:coefs} and Figure \ref{fig:var}, respectively. The average computational time for each simulation run was 9 seconds for \texttt{mice-default}, 68 seconds for \texttt{mice-cart}, 72 seconds for \texttt{mice-ranger}, 11 seconds for \texttt{mixgb}, and 12 seconds for \texttt{mixgb-sub}.

To evaluate the biases of the imputation methods, we compared the estimates of coefficients obtained from different MI implementations against the empirical true estimates and estimates found using complete cases. These estimates are displayed in Figure \ref{fig:coefs}. As can be seen, complete case estimates had the largest biases for most coefficients compared to MI. Among MI methods, \texttt{mice-default} performed worse than tree-based methods, especially for interaction terms. This is consistent with the results obtained by \cite{Doove2014}. Tree-based methods had rather similar performances. Out of 14 coefficients, \texttt{mixgb-sub} obtained 8 least biased estimates, \texttt{mixgb} obtained 3 least biased estimates and \texttt{mice-cart} also had 3. In this study, \texttt{mixgb-sub} had slightly smaller bias than \texttt{mixgb} for most coefficients, but the differences between these two methods were small.

We evaluated the performance of imputation variance, as discussed in Section \ref{sec:3.2}, by plotting the relative ratios $\text{Var}_W/\text{Var}_W^{\text{target}}-1$ and $\text{Var}_B/\text{Var}_B^{\text{target}}-1$ in Figure \ref{fig:var}. A relative ratio closer to zero indicates a better variance estimate. The top panel shows that all imputation methods obtained larger within-imputation variance than $\text{Var}_W^\text{target}$. This was expected as we generated a large proportion of missing values in most variables in the dataset. Additionally, $\texttt{mice-default}$ had larger within-imputation variances than other methods, followed by $\texttt{mice-ranger}$, while $\texttt{mice-cart}$ tended to have the smallest within-imputation variances.

In terms of between-imputation variance, we can see that \texttt{mixgb} without subsampling was lower than the target variance for all coefficients. This confirms our expectation that when subsampling is not used, mixgb will underestimate the between-imputation variance because it does not incorporate the uncertainty of model parameters. With subsampling, \texttt{mixgb-sub} was closer to the target variance, and it had the best performance overall. On the other hand, \texttt{mice-default}, \texttt{mice-cart}, and \texttt{mice-ranger} obtained higher between-imputation variances than the target values. Among these, \texttt{mice-ranger} had the largest relative ratio in 11 of 14 estimates, indicating that it might have overestimated the between-imputation variances.

As can be seen in Table \ref{tab:simtable}, all methods had zero coverage for terms \texttt{norm5}, \texttt{norm7}, \texttt{ord12}, \texttt{norm5:bin11}, and \texttt{norm7:ord11}. This was not surprising given our simulation settings. The coverage rates of the coefficients \texttt{norm1}, \texttt{norm2} and \texttt{norm3} were higher than those of other coefficients for all methods. Overall, \texttt{mice-default} had the lowest coverage rates for all interaction terms. Tree-based methods had similar coverage rates for most coefficients but \texttt{mice-cart} and \texttt{mice-ranger} performed considerably worse than \texttt{mixgb} and \texttt{mixgb-sub} for \texttt{bin11} and \texttt{ord11}. The coverage rates of \texttt{mixgb-sub} were slightly better than those of \texttt{mixgb}.

%Since the table is too large, we use a newpage and \setlength{\footskip}{70pt} to move the page number below the table caption. 
\newpage
\setlength{\footskip}{70pt}

\begin{table}[H]
	\centering
	\LARGE
	\renewcommand{\arraystretch}{0.6}
	\resizebox{0.65\textheight}{!}{%
		\begin{tabular}{rlcccccccc}
			\hline
			\rule{0pt}{4ex}
			& Method & \begin{tabular}[c]{@{}c@{}}Bias\\ \end{tabular} & \begin{tabular}[c]{@{}c@{}} $\text{Var}_T$ \\ $\times 1000$\end{tabular} & \begin{tabular}[c]{@{}c@{}}$\text{Var}_T^{\text{target}}$ \\ $\times 1000$\end{tabular}& \begin{tabular}[c]{@{}c@{}}$\text{Var}_W$\\ $\times 1000$\end{tabular}& \begin{tabular}[c]{@{}c@{}}$\text{Var}_W^{\text{target}}$\\ $\times 1000$\end{tabular}& \begin{tabular}[c]{@{}c@{}}$\text{Var}_B$\\ $\times 1000$\end{tabular}& \begin{tabular}[c]{@{}c@{}}$\text{Var}_B^{\text{target}}$\\ $\times 1000$\end{tabular} & CI coverage\%   \\ 
		\hline
			(Intercept) & mice-default & -0.29 & 12.32 & 6.00 & 5.98 & 1.11 & 6.35 & 4.89 & 10  \\ 
		 & mice-cart & -0.29 & 8.76 & 4.67 & 3.56 & 1.11 & 5.20 & 3.56 & 5 \\ 
		 & mice-ranger & -0.22 & 10.18 & 4.43 & 4.82 & 1.11 & 5.36 & 3.32 & 20  \\ 
		 & mixgb & -0.19 & 7.81 & 6.91 & 4.30 & 1.11 & 3.51 & 5.80 & 20  \\ 
		 & mixgb-sub & -0.20 & 8.68 & 7.31 & 4.21 & 1.11 & 4.46 & 6.20 & 25  \\ 
		\hline
		norm1 & mice-default & -0.01 & 2.88 & 1.85 & 1.05 & 0.18 & 1.83 & 1.67 & 91 \\ 
		 & mice-cart & -0.00 & 1.70 & 1.23 & 0.58 & 0.18 & 1.11 & 1.05 & 90   \\ 
		 & mice-ranger & -0.01 & 2.09 & 1.05 & 0.74 & 0.18 & 1.34 & 0.87 & 93  \\ 
		 & mixgb & -0.03 & 1.35 & 1.34 & 0.59 & 0.18 & 0.76 & 1.16 & 76  \\ 
		 & mixgb-sub & -0.02 & 1.46 & 1.32 & 0.58 & 0.18 & 0.88 & 1.14 & 81  \\ 
		\hline
		norm2 & mice-default & -0.03 & 2.66 & 1.76 & 1.04 & 0.18 & 1.62 & 1.58 & 86 \\ 
		 & mice-cart & -0.04 & 1.73 & 1.13 & 0.59 & 0.18 & 1.14 & 0.95 & 72  \\ 
		 & mice-ranger & -0.04 & 2.08 & 1.04 & 0.75 & 0.18 & 1.33 & 0.86 & 80  \\ 
		 & mixgb & -0.02 & 1.27 & 1.21 & 0.59 & 0.18 & 0.68 & 1.03 & 80  \\ 
		 & mixgb-sub & -0.01 & 1.40 & 1.17 & 0.58 & 0.18 & 0.82 & 0.99 & 85  \\ 
		\hline
		norm3 & mice-default & -0.06 & 2.65 & 1.65 & 1.05 & 0.18 & 1.59 & 1.47 & 61  \\ 
			 & mice-cart & -0.04 & 1.81 & 1.08 & 0.59 & 0.18 & 1.22 & 0.89 & 76  \\ 
			 & mice-ranger & -0.05 & 2.09 & 0.97 & 0.75 & 0.18 & 1.34 & 0.79 & 74  \\ 
			 & mixgb & -0.03 & 1.28 & 1.19 & 0.59 & 0.18 & 0.69 & 1.01 & 72  \\ 
			 & mixgb-sub & -0.02 & 1.45 & 1.14 & 0.58 & 0.18 & 0.87 & 0.96 & 79  \\ 
			\hline
		norm5 & mice-default & -0.71 & 3.65 & 1.67 & 1.74 & 0.31 & 1.91 & 1.37 & 0  \\ 
			 & mice-cart & -0.31 & 2.70 & 1.39 & 0.99 & 0.31 & 1.71 & 1.08 & 0  \\ 
			 & mice-ranger & -0.34 & 3.06 & 1.25 & 1.37 & 0.31 & 1.69 & 0.94 & 0  \\ 
			 & mixgb & -0.35 & 2.02 & 1.68 & 1.03 & 0.31 & 0.98 & 1.38 & 0  \\ 
			 & mixgb-sub & -0.35 & 2.30 & 1.89 & 1.03 & 0.31 & 1.27 & 1.59 & 0  \\ 
			\hline
		norm7 & mice-default & -0.49 & 6.96 & 2.63 & 3.46 & 0.61 & 3.50 & 2.02 & 0  \\ 
			 & mice-cart & -0.45 & 4.81 & 2.42 & 1.98 & 0.61 & 2.83 & 1.81 & 0  \\ 
			 & mice-ranger & -0.46 & 5.89 & 2.29 & 2.67 & 0.61 & 3.22 & 1.68 & 0  \\ 
			 & mixgb & -0.33 & 4.20 & 2.93 & 2.26 & 0.61 & 1.94 & 2.32 & 0  \\ 
			 & mixgb-sub & -0.33 & 4.58 & 3.02 & 2.23 & 0.61 & 2.35 & 2.41 & 0  \\ 
			\hline
		bin11 & mice-default & -0.12 & 8.48 & 4.65 & 3.24 & 0.59 & 5.24 & 4.07 & 64  \\ 
		 & mice-cart & -0.37 & 5.32 & 2.44 & 1.91 & 0.59 & 3.41 & 1.85 & 0  \\ 
		 & mice-ranger & -0.35 & 6.08 & 2.42 & 2.61 & 0.59 & 3.47 & 1.84 & 0  \\ 
		 & mixgb & -0.08 & 3.96 & 4.38 & 1.91 & 0.59 & 2.05 & 3.79 & 53  \\ 
		 & mixgb-sub & -0.08 & 4.63 & 4.32 & 1.90 & 0.59 & 2.73 & 3.74 & 64  \\ 
		\hline
		ord11 & mice-default & 0.42 & 11.89 & 6.11 & 5.77 & 1.07 & 6.13 & 5.04 & 0  \\ 
		 & mice-cart & 0.43 & 8.03 & 3.91 & 3.41 & 1.07 & 4.62 & 2.84 & 0  \\ 
		 & mice-ranger & 0.41 & 9.58 & 4.13 & 4.56 & 1.07 & 5.02 & 3.07 & 0  \\ 
		 & mixgb & 0.21 & 7.36 & 6.31 & 4.03 & 1.07 & 3.33 & 5.24 & 15  \\ 
		 & mixgb-sub & 0.20 & 8.21 & 6.72 & 3.96 & 1.07 & 4.26 & 5.65 & 22  \\ 
		\hline
		ord12 & mice-default & 1.25 & 22.48 & 13.88 & 9.75 & 1.74 & 12.73 & 12.13 & 0  \\ 
		 & mice-cart & 0.95 & 14.38 & 7.25 & 5.57 & 1.74 & 8.81 & 5.50 & 0  \\ 
		 & mice-ranger & 1.04 & 17.98 & 7.48 & 7.78 & 1.74 & 10.20 & 5.73 & 0  \\ 
		 & mixgb & 0.72 & 13.30 & 12.70 & 7.07 & 1.74 & 6.24 & 10.95 & 0  \\ 
		 & mixgb-sub & 0.74 & 15.06 & 13.96 & 6.89 & 1.74 & 8.17 & 12.21 & 0  \\ 
		\hline	
		I$(\text{norm1}^2)$ & mice-default & -0.34 & 0.96 & 0.44 & 0.35 & 0.06 & 0.60 & 0.37 & 0  \\ 
		 & mice-cart & -0.11 & 0.72 & 0.51 & 0.21 & 0.06 & 0.51 & 0.44 & 2  \\ 
		& mice-ranger & -0.14 & 1.04 & 0.43 & 0.27 & 0.06 & 0.77 & 0.36 & 1  \\ 
		 & mixgb & -0.14 & 0.56 & 0.59 & 0.21 & 0.06 & 0.35 & 0.53 & 0  \\ 
		 & mixgb-sub & -0.14 & 0.64 & 0.50 & 0.21 & 0.06 & 0.43 & 0.44 & 0  \\ 
		\hline
		norm2:norm3 & mice-default & -0.23 & 1.25 & 0.62 & 0.50 & 0.09 & 0.75 & 0.53 & 0  \\ 
		 & mice-cart & -0.11 & 1.09 & 0.84 & 0.29 & 0.09 & 0.80 & 0.76 & 12 \\ 
		 & mice-ranger & -0.13 & 1.40 & 0.68 & 0.38 & 0.09 & 1.01 & 0.59 & 8  \\ 
		 & mixgb & -0.11 & 0.75 & 0.82 & 0.29 & 0.09 & 0.46 & 0.73 & 4  \\ 
		 & mixgb-sub & -0.11 & 0.85 & 0.72 & 0.29 & 0.09 & 0.57 & 0.63 & 6  \\ 
		\hline
		norm5:bin11 & mice-default & 1.45 & 6.45 & 1.77 & 3.40 & 0.59 & 3.05 & 1.18 & 0  \\ 
		 & mice-cart & 0.92 & 5.85 & 3.07 & 1.93 & 0.59 & 3.92 & 2.48 & 0  \\ 
		 & mice-ranger & 0.93 & 6.68 & 2.43 & 2.67 & 0.59 & 4.01 & 1.84 & 0  \\ 
		 & mixgb & 0.77 & 3.79 & 3.49 & 1.95 & 0.59 & 1.83 & 2.90 & 0  \\ 
		 & mixgb-sub & 0.78 & 4.34 & 3.34 & 1.94 & 0.59 & 2.40 & 2.76 & 0  \\ 
		\hline
		norm7:ord11 & mice-default & 1.24 & 10.00 & 2.81 & 5.36 & 0.95 & 4.64 & 1.86 & 0 \\ 
		 & mice-cart & 0.96 & 8.23 & 4.39 & 3.08 & 0.95 & 5.16 & 3.44 & 0  \\ 
		 & mice-ranger & 0.96 & 10.03 & 3.98 & 4.14 & 0.95 & 5.89 & 3.02 & 0  \\ 
		 & mixgb & 0.80 & 6.50 & 4.85 & 3.33 & 0.95 & 3.17 & 3.90 & 0  \\ 
		 & mixgb-sub & 0.79 & 6.92 & 4.80 & 3.30 & 0.95 & 3.62 & 3.84 & 0  \\ 
		\hline
		norm7:ord12 & mice-default & -0.49 & 14.00 & 4.11 & 7.31 & 1.25 & 6.70 & 2.86 & 0  \\ 
		 & mice-cart & -0.29 & 11.04 & 5.99 & 4.04 & 1.25 & 7.00 & 4.74 & 14  \\ 
		 & mice-ranger & -0.33 & 14.10 & 5.66 & 5.59 & 1.25 & 8.51 & 4.41 & 9  \\ 
		 & mixgb & -0.28 & 9.40 & 6.68 & 4.63 & 1.25 & 4.77 & 5.43 & 8  \\ 
		 & mixgb-sub & -0.28 & 9.88 & 6.60 & 4.57 & 1.25 & 5.31 & 5.35 & 9  \\ 
		\hline
			&Maximum Monte Carlo SE &0.003  & ---  &---  &--- &---&---  &---  &2      \\
			&Maximum Bootstrap SE (100 samples) &---  &0.03  &0.05 &0.001 &0 & 0.03 &0.05  &---   \\
			\hline		
		\end{tabular}%
	}
	\caption{Summary of simulation results. {\scriptsize Note: $\text{Var}_T=\text{Var}_W+\text{Var}_B$,  $\text{Var}^{\text{target}}_T=\text{Var}^{\text{target}}_W+\text{Var}^{\text{target}}_B$.}} 
\label{tab:simtable}
\end{table}

\section{Computational Time}\label{sec:4}
%The previous page has changed the setlength to 70pt so that the page number is below the table caption. Here we change it back to the default setlength setting 30pt.
\setlength{\footskip}{30pt}

This section compares the computational time for different MI implementations using real and simulated datasets. As both XGBoost and \textbf{ranger} provide multithreading, we utilized all 20 available threads for \texttt{mixgb-cpu}, \texttt{mixgbsub-cpu} and \texttt{mice-ranger}. For \texttt{mixgb-gpu} and \texttt{mixgbsub-gpu}, only a single CPU thread was used with GPU support. Since XGBoost with GPU uses tree method \texttt{gpu\_hist}, we used tree method \texttt{hist} for the CPU version in order to eliminate alternative explanations for the difference in computational time between them. The default number of iterations \texttt{maxit} is five for the \textbf{mice} package, whereas in our package \textbf{mixgb}, the default \texttt{maxit} is one. In this comparison, we set $\texttt{maxit}$ to one for all methods. In practice, \textbf{mice} uses a higher value of \texttt{maxit} and it will be slower. Apart from these, we used default settings for all methods. All code used in this section is available in the supplementary materials.

 First, we chose  three real-world datasets of varying sizes to gain a sense of the imputation time for different MI implementations. For each dataset,  30\% missing data was created under the MCAR mechanism in two variables. Table \ref{tab:datatime} displays the average elapsed time to obtain five imputed datasets over ten replications.

\begin{table}[h]
	\centering
		\resizebox{0.65\textwidth}{!}{%
	\begin{tabular}{lcccccc}
		\hline
		& \multicolumn{3}{c}{\textbf{Dataset}}                   \\ \cline{2-7} 
		\textbf{Methods}      & \href{https://archive.ics.uci.edu/ml/datasets/default+of+credit+card+clients}{\textbf{Credit}} & \textbf{s.e} & \href{https://www.kaggle.com/competitions/allstate-claims-severity/data}{\textbf{Allstate}}& \textbf{s.e} &\href{https://archive.ics.uci.edu/ml/datasets/HIGGS}{\textbf{Higgs1M}} & \textbf{s.e}\\ \hline
		\textbf{mice-default} & 3.7   &  (0.03)  & 2104.7 & (11.5)      & 112.2   & (4.8)         \\
		\textbf{mice-cart}    & 18.7  & (0.1)    & 3651.9 & (10.0)       & 3587.0  & (5.2)       \\
		\textbf{mice-ranger}  & 8.2   & (0.02)   & 537.9 & (1.4)        & 397.3  & (10.8)  \\
		\textbf{mixgb-cpu}    & 3.0  & (0.03)      &62.8   & (0.7)        & 98.7  & (4.5)       \\
		\textbf{mixgb-gpu}    & 3.9  & (0.09)     &60.7   & (0.5)        & 55.6  & (2.0)          \\ 
		\textbf{mixgbsub-cpu}   & 3.7  & (0.03)   &66.0   &  (0.7)       & 125.9 & (4.1)        \\
		\textbf{mixgbsub-gpu}   & 4.0 & (0.06)    &60.9   & (1.1)        & 56.5 & (1.9)          \\ 
		\hline
	\end{tabular} 
}
	\caption{Average computational time and standard error (s.e) for imputing real datasets using various MI implementations. Results are averaged over 10 repetitions.}\label{tab:datatime}
\end{table}
 
The dataset \textbf{Credit} \citep{Yeh2009} was used to predict credit card defaults in Taiwan and can be obtained from the UCI machine learning repository \citep{Kelly2023}. It has 30000 samples and 24 features (20 continuous and 4 categorical features). The dataset \textbf{Allstate} was used in a famous Kaggle competition ``Allstate Claims Severity" \citep{Kaggle2016}. The training set has 188318 samples and 131 features (15 continuous and 116 categorical features). The dataset \textbf{Higgs1M} \citep{Baldi2014} has 1 million samples and 28 features (27 continuous features and 1 categorical feature). The original dataset has 11 million samples and can also be obtained from the UCI repository \citep{Kelly2023}. In this experiment, we only took the first 1 million samples.

We can see that for the datasets \textbf{Credit} and \textbf{Higgs1M}, which have mainly continuous features, \texttt{mice-default} was quite fast. However, it was much slower to impute \textbf{Allstate}, which has more categorical features. On the other hand, \texttt{mice-cart} was the slowest for all datasets. Compared to \texttt{mice-cart}, \texttt{mice-ranger} was much faster, yet it was slower than different versions of \textbf{mixgb}. With GPU support, \texttt{mixgb-gpu} and \texttt{mixgbsub-gpu} were faster than their CPU counterparts \texttt{mixgb-cpu} and \texttt{mixgbsub-cpu} for imputing large datasets \textbf{Allstate} and \textbf{Higgs1M}, but CPU versions were faster for imputing the smaller dataset \textbf{Credit}. Subsampling with \textbf{mixgb} took more time when just the CPU was used. However, subsampling required significantly less time when the GPU was utilised.

To further investigate the computational time of different MI implementations, we ran an experiment on simulated data. We generated datasets with sample sizes ranging from $10^3$ to $10^6$, with the number of features being 11, 31, and 51. Standard normal data were created for continuous data. For categorical data, all explanatory variables had three levels and the response variable was binary. For each dataset, 50\% missing data in three variables were created via the MCAR mechanism. We did not evaluate \texttt{mice-cart} here as it would have required excessive computation time even for medium-sized datasets. 

\begin{figure}[H]
	\centering
	\includegraphics[width=\textwidth,keepaspectratio]{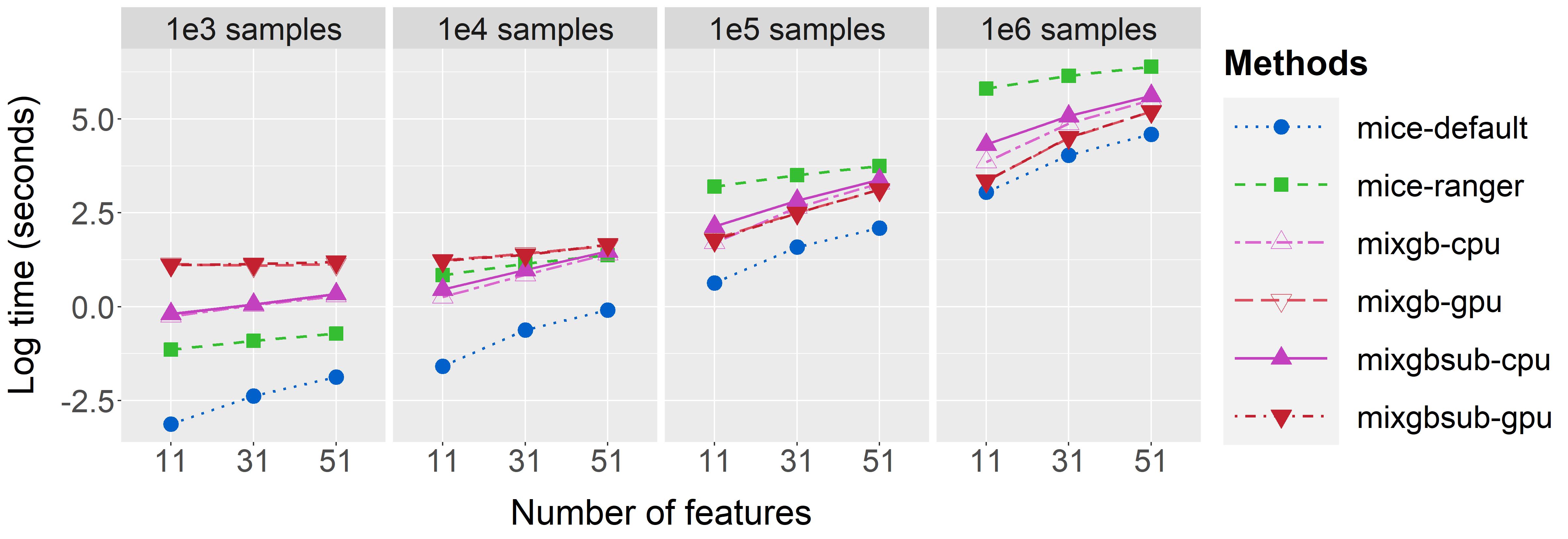}
	\caption{Computational time for imputing continuous data.}\label{fig:timecon}
\end{figure}

\begin{figure}[H]
	\centering
	\includegraphics[width=\textwidth,keepaspectratio]{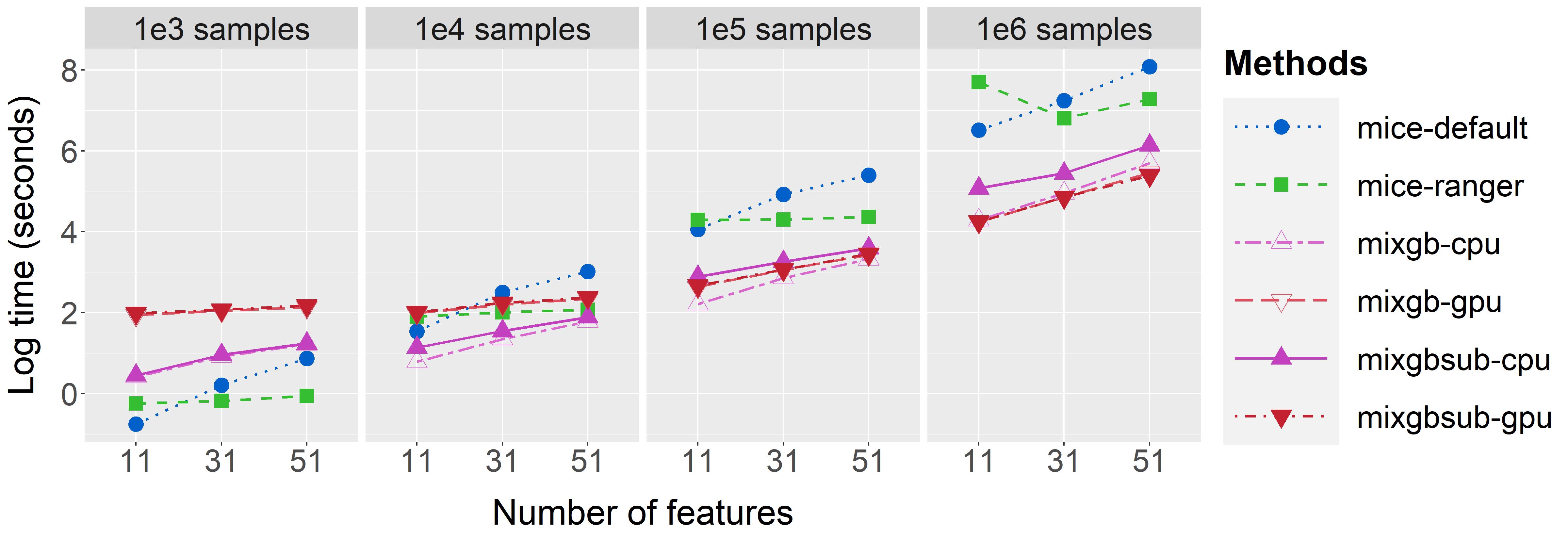}
	\caption{Computational time for imputing categorical data.}\label{fig:timecat}
\end{figure}

The log average time to obtain five imputations for continuous data and categorical data is shown in Figures \ref{fig:timecon} and \ref{fig:timecat}, respectively, with all results averaged over ten repetitions. Overall, \texttt{mice-default} was the fastest for continuous datasets, but it performed poorly with increasing numbers of observations and features for categorical data. On the other hand, \texttt{mice-ranger} was reasonably fast for datasets with relatively small sample sizes; however, as the sample size increased, it scaled badly for both continuous and categorical data. To a certain extent, \texttt{mice-ranger} handled larger numbers of features more effectively than larger number of samples. Surprisingly, when working with the $10^6\times 11$ dataset, \texttt{mice-ranger} took considerably longer to run compared to other methods. We will discuss this outlier later. All four versions of \textbf{mixgb} had good scalability across varying numbers of features and samples, and outperformed \texttt{mice-default} and \texttt{mice-ranger} for large categorical datasets, where the GPU versions of \textbf{mixgb} were the fastest methods. When using multithreading CPU, \texttt{mixgb-cpu} and \texttt{mixgbsub-cpu} were faster than their GPU counterpart for smaller datasets, but \textbf{mixgb} with GPU support performed better for large sample sizes. Our experiment also showed that using subsampling with CPU increased imputation time, especially when the sample size was large. However, the additional time required for subsampling was minor with GPU support.

The computational time for \texttt{mice-ranger} on the dataset with $10^6$ observations and 11 features may seem counter-intuitive. It was the only place where the runtime was larger with fewer features. We profiled the \texttt{mice-ranger} imputation process for the three datasets with $10^6$ observations. We found that the time taken to fit the imputation model using ranger increased with the number of features, as expected, and the resulting tree for 11 features was simpler and shallower. However, the number of observations at each terminal node was much larger. When \texttt{mice-ranger} obtains predicted values from a given tree, it randomly samples one observation from the corresponding terminal node. For the dataset with 11 features, each tree had approximately $300$ times as many observations at each terminal node on average, compared to the datasets with 31 and 51 features, which had similar numbers of observations at each terminal node. Since all datasets had $5\times 10^5$ missing data points in each missing variable, the increase in time required to sample from terminal nodes was significant and outweighed the shorter time to build the tree. This effect caused the unusual time.

\section{Data Example}\label{sec:5}
The NWTS (US National Wilms' Tumor Study) dataset \citep{DAngio1989} contains outcome variables (e.g., \texttt{relaps} and \texttt{dead}) as well as histology results for 3915 patients. Since it provides both the central lab histology (\texttt{histol}) and local institution histology (\texttt{instit}) results, it has been extensively used as an example to illustrate two-phase sampling designs \citep[e.g.,][]{Breslow1999,Breslow2009,Chen2020}. Other variables include \texttt{study}  (3 or 4), \texttt{stage} (I-IV), the \texttt{age} and \texttt{year} at diagnosis, the weight of tumor (\texttt{specwgt}), the diameter of tumor (\texttt{tumdiam}), time to relapse (\texttt{trel}) and time to death (\texttt{tsur}). There are no missing values in the original dataset. We created missing data  in \texttt{histol}, \texttt{tumdiam} and \texttt{stage} by an MAR mechanism dependent on \texttt{relaps}, \texttt{instit}, and \texttt{specwgt}. Missing data in \texttt{histol} and \texttt{stage} depended on \texttt{relaps} and \texttt{instit}; missing data in \texttt{tumdiam} depended on \texttt{relaps} and \texttt{specwgt}. Full details can be found in the supplementary materials. Five imputed datasets were then generated using \texttt{mice-default}, \texttt{mice-cart}, \texttt{mice-ranger}, \texttt{mixgb} and \texttt{mixgb-sub}. We fitted the following analysis model based on \cite{Kulich2004}:
\begin{align}	
h(t) &= h_0(t) \times \exp\left(\beta_1 \text{histol}+\beta_2\text{age}_1+\beta_3 \text{age}_2+\beta_4\text{stage}+\beta_5\text{tumdiam}\right. \nonumber\\
&\left.+\beta_6\text{histol}\times \text{age}_1+\beta_7\text{histol}\times \text{age}_2+\beta_8\text{stage}\times\text{tumdiam}\right),
\end{align}

where $\beta_2$ and $\beta_3$ are separate slopes for age younger than one year ($\texttt{age}_1$) and age one year or older ($\texttt{age}_2$). The variable \texttt{stage} is a binary indicator. It is coded 1 to indicate stage III-IV and 0 to indicate stage I-II. Table \ref{tab:nwtscoefs} summarizes the coefficient estimates obtained from full data, complete cases and MI methods. All MI implementations produced better estimates than complete case analysis.

\begin{table}[H]
	\centering
	\resizebox{\textwidth}{!}{%
		\begin{tabular}{lcccccccc}
			\hline
			\multicolumn{1}{c}{\multirow{2}{*}{\textbf{Estimates}}} &  & \multicolumn{7}{c}{\textbf{Methods}}                                                                                           \\ \cline{3-9}
			\multicolumn{1}{c}{}                                    &  & \textbf{full data} &\textbf{complete cases}& \textbf{mice-default} & \textbf{mice-cart} & \textbf{mice-ranger}  & \textbf{mixgb} &\textbf{mixgb-sub}\\ \hline
			$\beta_1$                                               &  & 4.10 & 6.75 & 4.10 & 4.32 & 4.39 & 4.19 & 4.26 \\
			$\beta_2$                                               &  &-0.66 & -1.21 & -0.67 & -0.59 & -0.62 & -0.63 & -0.58\\
			$\beta_3$                                               &  &0.10 & 0.14 & 0.10 & 0.10 & 0.10 & 0.10 & 0.10\\
			$\beta_4$                                               &  &1.35 & 2.59 & 1.20 & 1.46 & 1.08 & 1.10 & 1.10\\
			$\beta_5$                                               &  &0.07 & 0.17 & 0.07 & 0.07 & 0.08 & 0.05 & 0.06\\
			$\beta_6$                                               &  &-2.64 & -4.59 & -2.65 & -2.95 & -2.98 & -2.71 & -2.79\\
			$\beta_7$                                               &  &-0.06 & -0.06 & -0.04 & -0.04 & -0.05 & -0.04 & -0.04\\
			$\beta_8$                                               &  & -0.08 & -0.13 & -0.07 & -0.09 & -0.06 & -0.05 & -0.06\\  \hline
		\end{tabular}%
	}
	\caption{Coefficient estimates of NWTS data using full data, complete cases and MI methods.}\label{tab:nwtscoefs}
\end{table}

\begin{figure}[H]
	\centering
	\includegraphics[width=\textwidth,keepaspectratio]{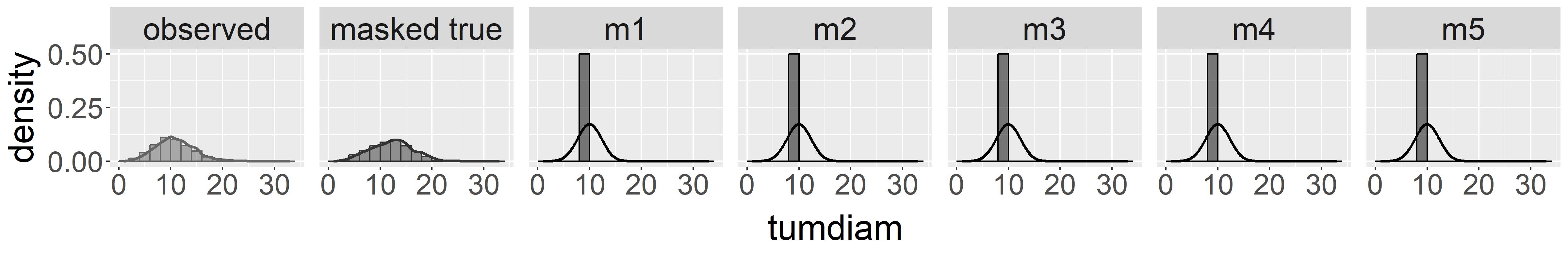}
	\caption{The comparison of the marginal distributions of the observed values, the masked true values, and five sets of imputed values using mean imputation for variable \texttt{tumdiam}.}\label{fig:imputemean}
\end{figure}

\begin{figure}[H]
	\centering
	\includegraphics[width=\textwidth,keepaspectratio]{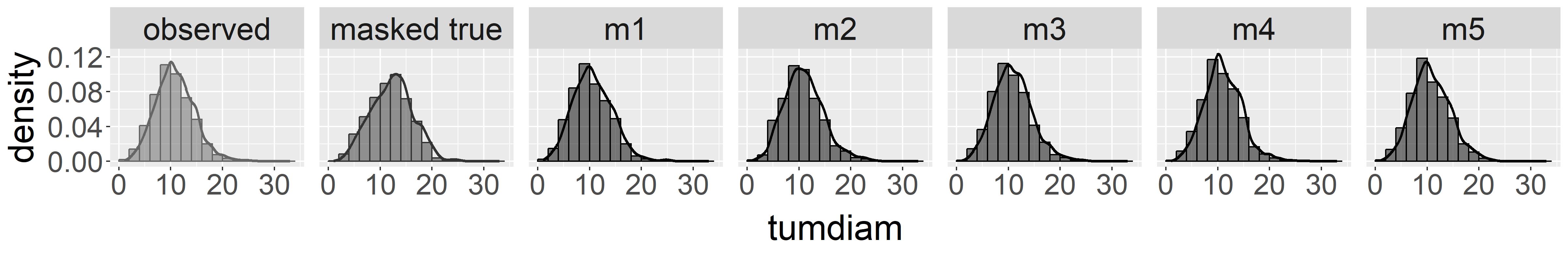}
	\includegraphics[width=\textwidth,keepaspectratio]{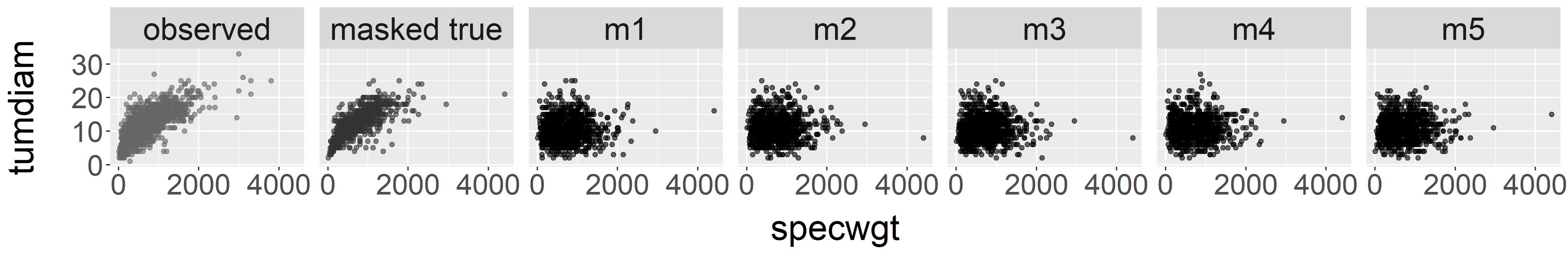}
	\caption{Multiply-imputed values using random sampling imputation. The top panel shows the marginal distributions for variable \texttt{tumdiam}; the bottom panel shows the joint distribution of variables \texttt{tumdiam} and \texttt{specwgt}.}\label{fig:imputesample}
\end{figure}

In general, we know neither the missing data mechanism nor the true values of the missing data. The distribution of imputed values can differ noticeably from that of observed values. However, we can resort to diagnostic tools to check whether the imputed values are plausible and whether the imputation variability is reasonable. Our \texttt{R} package \textbf{mixgb} includes some functions for visualizing multiply-imputed values. Figure \ref{fig:imputemean} compares the marginal distribution of the observed data, the masked true data and multiply-imputed data using mean imputation in the variable \texttt{tumdiam}. In this case, there was no variability among five imputations. If we impute missing values by randomly sampling from the observed values of a variable, then the marginal distribution of that variable will resemble that of the observed values, like what is shown in the top panel of Figure \ref{fig:imputesample}. The bottom panel of Figure \ref{fig:imputesample}, however, illustrates that random sampling imputation failed to capture the relationship between variables \texttt{tumdiam} and \texttt{specwgt}.

\begin{figure}[h]
	\centering
	\includegraphics[width=\textwidth,keepaspectratio]{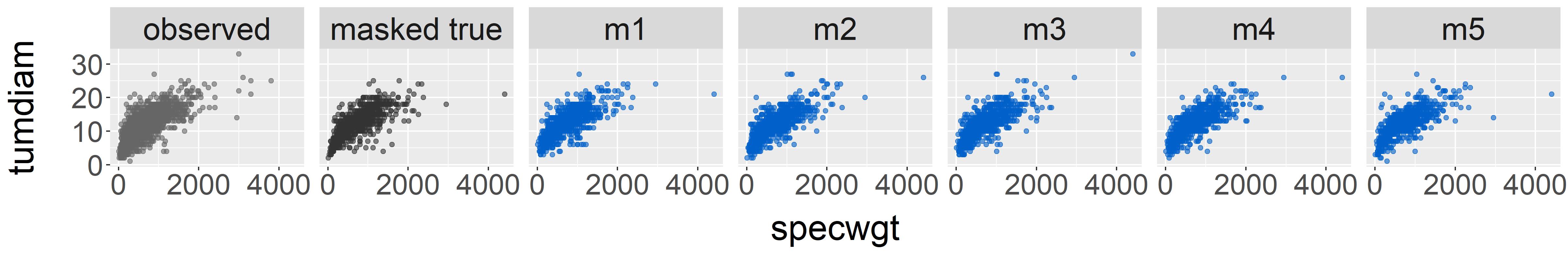}
	\includegraphics[width=\textwidth,keepaspectratio]{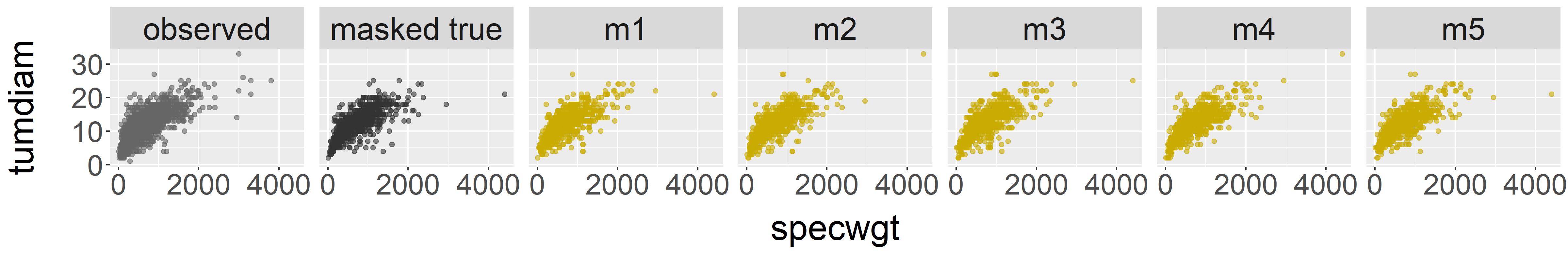}
	\includegraphics[width=\textwidth,keepaspectratio]{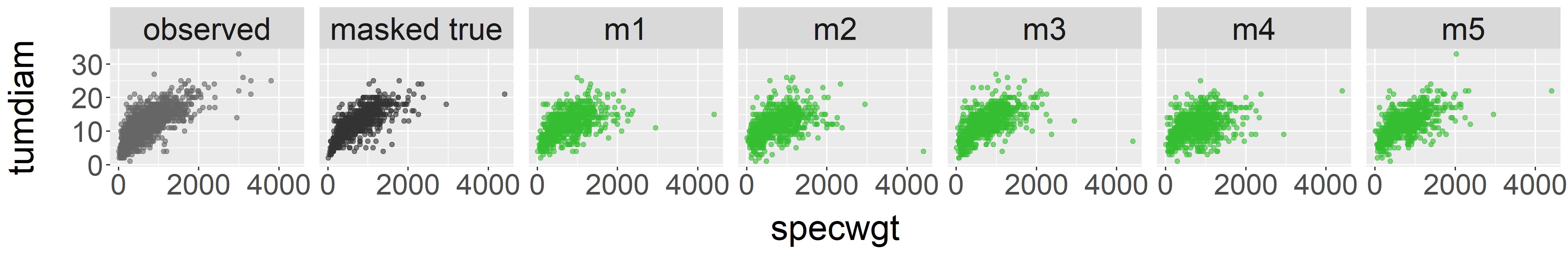}
	\includegraphics[width=\textwidth,keepaspectratio]{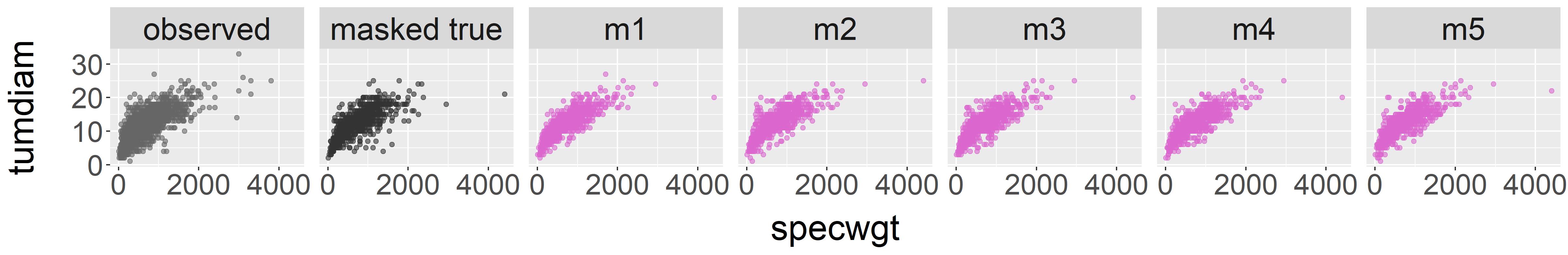}
    \includegraphics[width=\textwidth,keepaspectratio]{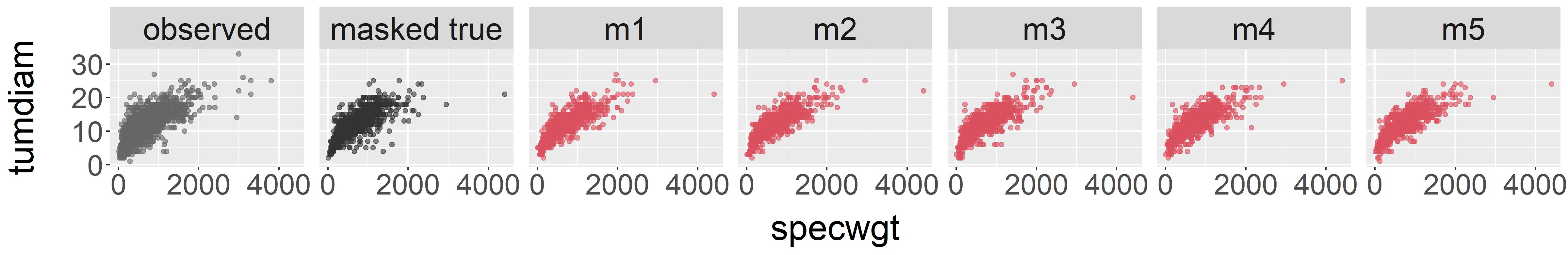}
	\caption{Multiply-imputed missing values using MI methods. From top to bottom: \texttt{mice-default}, \texttt{mice-cart}, \texttt{mice-ranger}, \texttt{mixgb}, \texttt{mixgb-sub}.}\label{fig:scatter}
\end{figure}

On the other hand, Figure \ref{fig:scatter} shows that using MI methods can preserve the relationship between variables. The plot also indicates that \texttt{mice-ranger} had larger within and between imputation variability. Other types of diagnostic plots, which are omitted here, show similar patterns across all methods.

\section{Discussion}\label{sec:6}
Standard MI implementations, such as \texttt{mice-default}, do not automatically account for interaction effects among variables in data. Applying tree-based algorithms such as CART and random forests to MI has been shown to alleviate this problem \citep{Doove2014}. However, tree-based algorithms can be computationally demanding for large datasets. 

In this study, we presented the design and imputation performance of \textbf{mixgb}, a MI implementation based on another tree-based algorithm XGBoost \citep{Chen2016}, subsampling, and predictive mean matching \citep{Little1988}. In our simulation study, we found that with subsampling, \texttt{mixgb-sub} achieved the least biased estimates on average compared to other MI implementations. Additionally, \texttt{mixgb-sub} performed the best in terms of between-imputation variance. We also demonstrated that all four versions of \textbf{mixgb} were slower for small datasets but scaled well for larger datasets, with GPU versions being the fastest. Subsampling increased the imputation time notably when utilizing the CPU, especially for large sample sizes. Nevertheless, this increase was negligible when the GPU was used.

To the best of our knowledge, this is the first attempt to implement and evaluate MI using XGBoost with predictive mean matching and subsampling. We show that \texttt{mixgb} and \texttt{mixgb-sub} can achieve imputation performances as good as that of \texttt{mice-cart}, which was previously regarded as the best method for capturing complex data structures \citep{Doove2014}. We consider \texttt{mice-ranger}, \texttt{mixgb} and \texttt{mixgb-sub} to be better alternatives for MI, since CART is susceptible to overfitting and does not generalize well in real-world data scenarios. Moreover, using \texttt{mice-cart} for large-scale imputations can be time-consuming. For smaller datasets, \texttt{mice-ranger} was faster than \texttt{mixgb-cpu} or \texttt{mixgb-gpu}; however, it was significantly slower for larger datasets. In addition, \texttt{mice-ranger} can take an excessive amount of time to run under certain circumstances, such as the case we described in Section \ref{sec:4}. Our findings suggest that \textbf{mixgb} is a promising automated MI framework for large and complex datasets. In this study, We found \texttt{mixgbsub-gpu} to be the best among all evaluated MI methods due to its speed and imputation performance.

Our results may be limited to simulation studies similar to the one presented in this paper and may not generalize to other scenarios. Under our simulation settings, a large proportion of missing values were created via an MAR mechanism that made complete case analysis invalid. If the data is missing completely at random or missingness is only weakly associated with the available variables, then complete case analysis can often achieve better results than MI methods. It is also worth noting that the speed and quality of imputation generally depend on hyperparameter settings, such as the number of trees in \texttt{mice-ranger} (default 10), the number of boosting rounds in \texttt{mixgb} and \texttt{mixgb-sub} (default 100), and the number of threads used when a multicore processor is available. Future work should evaluate imputation performance with diverse datasets and a wider range of hyperparameter configurations.

\begin{center}
	{\large\bf Computation Details}
\end{center}
The simulation study in Section \ref{sec:3} on imputation performance was run using \texttt{R} 4.0.2 \citep{RCT2022} on a Ubuntu 18.04 server with a 64-core Intel Xeon Gold 6142 CPU processor and 187GB of RAM, without the use of GPU. Simulation tasks for all methods were run in parallel. The evaluation of computational time in Section \ref{sec:4} was run using \texttt{R} 4.2.1 \citep{RCT2022} on a 64-bit Windows 10 PC with 64 GB of RAM, an Intel i7-12700K CPU, and an Nvidia GeForce RTX 3080 Ti GPU. Each imputation method was tested separately to avoid interference between tasks. \texttt{R} package \textbf{mice} \citep{Buuren2011} version 3.14.0 and \texttt{R} package \textbf{mixgb} \citep{Deng2023} version 1.0.1 were used in this paper. Computational time was recorded using the \texttt{R} package \textbf{microbenchmark} \citep{Mersmann2021} and visualization of simulation results used the \texttt{R} package \textbf{ggplot2} \citep{Wickham2016}.

\begin{center}
	{\large\bf Supplementary Materials}
\end{center}
All supplementary materials are contained in a single archive \texttt{supplement.zip}. For more details, please refer to the \texttt{readme.md} file in the zip file.

\begin{center}
	{\large\bf Acknowledgements}
\end{center}
The authors would like to express their sincere gratitude to the Editor, Associate Editor and two Reviewers for their insightful comments and suggestions.

\begin{center}
	{\large\bf Disclosure Statement}
\end{center}
The authors report there are no competing interests to declare.

\bibliographystyle{JASA.bst}
\bibliography{mixgb}

\begin{thebibliography}{30}
\newcommand{\enquote}[1]{``#1''}
\expandafter\ifx\csname natexlab\endcsname\relax\def\natexlab#1{#1}\fi

\bibitem[{Awada et~al.(2021)Awada, Bouaoun, Nasr, Tfayli, Cuenin, Akika,
  Boustany, Makoukji, Tamim, Zgheib, and Ghantous}]{Awada2021}
Awada, Z., Bouaoun, L., Nasr, R., Tfayli, A., Cuenin, C., Akika, R., Boustany,
  R.-M., Makoukji, J., Tamim, H., Zgheib, N.~K.,  and Ghantous, A. (2021),
  \enquote{{LINE-1 Methylation Mediates the Inverse Association Between Body
  Mass Index and Breast Cancer Risk: A Pilot Study in the Lebanese
  Population},} \textit{Environmental Research}, 197, 111094.

\bibitem[{Baldi et~al.(2014)Baldi, Sadowski, and Whiteson}]{Baldi2014}
Baldi, P., Sadowski, P.,  and Whiteson, D. (2014), \enquote{Searching for
  Exotic Particles in High-energy Physics with Deep Learning,} \textit{Nature
  communications}, 5, 4308.

\bibitem[{Brand et~al.(2003)Brand, Van~Buuren, Groothuis-Oudshoorn, and
  Gelsema}]{Brand2003}
Brand, J.~P., Van~Buuren, S., Groothuis-Oudshoorn, K.,  and Gelsema, E.~S.
  (2003), \enquote{{A Toolkit in SAS for the Evaluation of Multiple Imputation
  Methods},} \textit{Statistica Neerlandica}, 57, 36--45.

\bibitem[{Breiman(2001)}]{Breiman2001}
Breiman, L. (2001), \enquote{{Random Forests},} \textit{Machine Learning}, 45,
  5--32.

\bibitem[{Breiman et~al.(1984)Breiman, Friedman, Olshen, and
  Stone}]{Breiman1984}
Breiman, L., Friedman, J.~H., Olshen, R.~A.,  and Stone, C.~J. (1984),
  \textit{{Classification and Regression Trees}}, Belmont, CA: Wadsworth.

\bibitem[{Breslow and Chatterjee(1999)}]{Breslow1999}
Breslow, N.~E.,  and Chatterjee, N. (1999), \enquote{{Design and Analysis of
  Two-Phase Studies with Binary Outcome Applied to Wilms Tumour Prognosis},}
  \textit{Journal of the Royal Statistical Society: Series C (Applied
  Statistics)}, 48, 457--468.

\bibitem[{Breslow et~al.(2009)Breslow, Lumley, Ballantyne, Chambless, and
  Kulich}]{Breslow2009}
Breslow, N.~E., Lumley, T., Ballantyne, C.~M., Chambless, L.~E.,  and Kulich,
  M. (2009), \enquote{{{Using the Whole Cohort in the Analysis of Case-Cohort
  Data}},} \textit{American Journal of Epidemiology}, 169, 1398--1405.

\bibitem[{Chen and Guestrin(2016)}]{Chen2016}
Chen, T.,  and Guestrin, C. (2016), \enquote{{XGBoost: A Scalable Tree Boosting
  System},} in \textit{Proceedings of the 22nd ACM SIGKDD International
  Conference on Knowledge Discovery and Data Mining}, New York, NY, USA:
  Association for Computing Machinery, pp. 785--794.

\bibitem[{Chen and Lumley(2020)}]{Chen2020}
Chen, T.,  and Lumley, T. (2020), \enquote{{Optimal Multiwave Sampling for
  Regression Modeling in Two-Phase Designs},} \textit{Statistics in Medicine},
  39, 4912--4921.

\bibitem[{D'Angio et~al.(1989)D'Angio, Breslow, Beckwith, Evans, Baum,
  Delorimier, Fernbach, Hrabovsky, Jones, Kelalis, et~al.}]{DAngio1989}
D'Angio, G.~J., Breslow, N., Beckwith, J.~B., Evans, A., Baum, E., Delorimier,
  A., Fernbach, D., Hrabovsky, E., Jones, B., Kelalis, P. et~al. (1989),
  \enquote{{Treatment of Wilms' Tumor. Results of the Third National Wilms'
  Tumor Study},} \textit{Cancer}, 64, 349--360.

\bibitem[{Deng(2023)}]{Deng2023}
Deng, Y. (2023), \textit{mixgb: Multiple Imputation Through XGBoost}, {R
  package version 1.0.1}.

\bibitem[{Doove et~al.(2014)Doove, van Buuren, and Dusseldorp}]{Doove2014}
Doove, L.~L., van Buuren, S.,  and Dusseldorp, E. (2014), \enquote{{Recursive
  Partitioning for Missing Data Imputation in the Presence of Interaction
  Effects},} \textit{Computational Statistics \& Data Analysis}, 72, 92--104.

\bibitem[{Kaggle(2016)}]{Kaggle2016}
Kaggle (2016), \textit{Allstate Claims Severity Dataset}, available at
  \url{https://www.kaggle.com/competitions/allstate-claims-severity/data}.

\bibitem[{Kelly et~al.(2023)Kelly, Longjohn, and Nottingham}]{Kelly2023}
Kelly, M., Longjohn, R.,  and Nottingham, K. (2023), \textit{{The UCI Machine
  Learning Repository}}, available at \url{https://archive.ics.uci.edu}.

\bibitem[{Kulich and Lin(2004)}]{Kulich2004}
Kulich, M.,  and Lin, D.~Y. (2004), \enquote{{Improving the Efficiency of
  Relative-Risk Estimation in Case-Cohort Studies},} \textit{Journal of the
  American Statistical Association}, 99, 832--844.

\bibitem[{Little(1988)}]{Little1988}
Little, R.~J. (1988), \enquote{{Missing-Data Adjustments in Large Surveys},}
  \textit{Journal of Business \& Economic Statistics}, 6, 287--296.

\bibitem[{Mersmann(2021)}]{Mersmann2021}
Mersmann, O. (2021), \textit{{microbenchmark: Accurate Timing Functions}}, {R
  package version 1.4.9}.

\bibitem[{{R Core Team}(2022)}]{RCT2022}
{R Core Team} (2022), \textit{{R: A Language and Environment for Statistical
  Computing}}, R Foundation for Statistical Computing, Vienna, Austria.

\bibitem[{Rubin(1978)}]{Rubin1978}
Rubin, D.~B. (1978), \enquote{{Multiple Imputations in Sample Surveys-A
  Phenomenological Bayesian Approach to Nonresponse},} in \textit{{Proceedings
  of the Survey Research Methods Section}}, American Statistical Association,
  pp. 20--34.

\bibitem[{Rubin(1986)}]{Rubin1986}
--- (1986), \enquote{{Statistical Matching using File Concatenation with
  Adjusted Weights and Multiple Imputations},} \textit{Journal of Business \&
  Economic Statistics}, 4, 87--94.

\bibitem[{Rubin(1987)}]{Rubin1987}
--- (1987), \textit{Multiple imputation for nonresponse in surveys}, Wiley
  Series in Probability and Mathematical Statistics: Applied Probability and
  Statistics, New York: John Wiley \& Sons, Inc.

\bibitem[{Stekhoven and B{\"u}hlmann(2012)}]{Stekhoven2012}
Stekhoven, D.~J.,  and B{\"u}hlmann, P. (2012),
  \enquote{{MissForest--Non-Parametric Missing Value Imputation for Mixed-Type
  Data},} \textit{Bioinformatics}, 28, 112--118.

\bibitem[{Su et~al.(2011)Su, Gelman, Hill, and Yajima}]{Su2011}
Su, Y.~S., Gelman, A., Hill, J.,  and Yajima, M. (2011), \enquote{{Multiple
  Imputation with Diagnostics (mi) in R: Opening Windows into the Black Box},}
  \textit{Journal of Statistical Software}, 45, 1--31.

\bibitem[{van Buuren(2018)}]{Buuren2018}
van Buuren, S. (2018), \textit{Flexible Imputation of Missing Data}, Boca
  Raton: Chapman \& Hall/CRC Press, 2nd ed.

\bibitem[{van Buuren and Groothuis-Oudshoorn(2011)}]{Buuren2011}
van Buuren, S.,  and Groothuis-Oudshoorn, K. (2011), \enquote{{Mice:
  Multivariate Imputation by Chained Equations in R},} \textit{Journal of
  Statistical Software}, 45, 1--67.

\bibitem[{Wendt et~al.(2021)Wendt, Pathak, Levey, Nuñez, Overstreet, Tyrrell,
  Adhikari, {De Angelis}, Tylee, Goswami, Krystal, Abdallah, Stein, Kranzler,
  Gelernter, and Polimanti}]{Wendt2021}
Wendt, F.~R., Pathak, G.~A., Levey, D.~F., Nuñez, Y.~Z., Overstreet, C.,
  Tyrrell, C., Adhikari, K., {De Angelis}, F., Tylee, D.~S., Goswami, A.,
  Krystal, J.~H., Abdallah, C.~G., Stein, M.~B., Kranzler, H.~R., Gelernter,
  J.,  and Polimanti, R. (2021), \enquote{Sex-{Stratified}
  {Gene}-by-{Environment} {Genome}-{Wide} {Interaction} {Study} of {Trauma},
  {Posttraumatic}-{Stress}, and {Suicidality},} \textit{Neurobiology of
  Stress}, 14, 100309.

\bibitem[{Wickham(2016)}]{Wickham2016}
Wickham, H. (2016), \textit{{ggplot2: Elegant Graphics for Data Analysis}}, New
  York: Springer-Verlag.

\bibitem[{Wright and Ziegler(2017)}]{Wright2017}
Wright, M.~N.,  and Ziegler, A. (2017), \enquote{{Ranger: A Fast Implementation
  of Random Forests for High Dimensional Data in C++ and R},} \textit{Journal
  of Statistical Software}, 77, 1--17.

\bibitem[{Yeh and Lien(2009)}]{Yeh2009}
Yeh, I.-C.,  and Lien, C.-h. (2009), \enquote{{The Comparisons of Data Mining
  Techniques for the Predictive Accuracy of Probability of Default of Credit
  Card Clients},} \textit{Expert systems with applications}, 36, 2473--2480.

\bibitem[{Zhang et~al.(2019)Zhang, Yan, Gao, Malin, and Chen}]{Zhang2019}
Zhang, X., Yan, C., Gao, C., Malin, B.,  and Chen, Y. (2019), \enquote{{XGBoost
  Imputation for Time Series Data},} in \textit{{2019 IEEE International
  Conference on Healthcare Informatics (ICHI)}}, pp. 1--3.

\end{thebibliography}
\end{document}